\documentclass[preprint2]{aastex}

\slugcomment{submitted to ApJ}

\shorttitle{X-Rays from M82} 
\shortauthors{Kaaret, Simet, \& Lang}

\begin{document}

\title{A 62 Day X-Ray Periodicity and an X-Ray Flare from the
Ultraluminous X-Ray Source in M82}

\author{Philip Kaaret, Melanie G.\ Simet, and Cornelia C.\ Lang} 
\affil{Department of Physics and Astronomy, University of Iowa,  Van
Allen Hall, Iowa City, IA 52242.}


\begin{abstract}

In 240 days of X-ray monitoring of M82, we have discovered an X-ray
periodicity at $62.0 \pm 2.5$ days with a peak to peak amplitude
corresponding to an isotropic luminosity of $2.4 \times 10^{40} \rm \,
erg \, s^{-1}$ in M82 and an X-ray flare reaching a peak luminosity of 
$9.8 \times 10^{40} \rm \, erg \, s^{-1}$.  The periodicity and flare
likely originate from the ultraluminous X-ray source (ULX) in M82 which
has been identified as a possible intermediate mass black hole.  We
suggest that the 62~day modulation is due to orbital motion within an
X-ray binary with a Roche-lobe overflowing companion star which would
imply that the average density of the companion star is near $5 \times
10^{-5} \rm \, g \, cm^{-3}$ and is therefore a giant or supergiant. 
Chandra observations just after the flare show an energy spectrum that
is consistent with a power-law with no evidence of a thermal component
or line emission.  Radio observations made with the VLA during the
flare allow us to rule out a blazar identification for the source and
place strong constraints on relativistically beamed models of the X-ray
emission.  The Chandra observations reveal a second X-ray source
reached a flux of $4.4 \times 10^{-12} \rm \, erg \, cm^{-2} \, s^{-1}$
in the 0.3-7~keV band which is dramatically higher than any flux
previously seen from this source and corresponds to an isotropic
luminosity of $1.1 \times 10^{40} \rm \, erg \, s^{-1}$.  This source
is a second ultraluminous X-ray source in M82 and may give rise to the
QPOs detected from the central region of M82.

\end{abstract}


\section{Introduction}

The bright X-ray sources in external galaxies, known as ultraluminous
X-ray sources (ULXs), are of considerable current interest because they
may be `intermediate-mass' black holes
\citep{colbert99,makishima00,kaaret01}.  One of the most extreme ULXs
is located in the nearby starburst galaxy M82 \citep{ptak99}.  Chandra
observations showed that the brightest source in the galaxy, CXOU
J095550.2+694047 =  X41.4+60 \citep{kaaret01,matsumoto01}, is offset
from the nucleus, not coincident with any radio supernova remnant, and
highly variable on time scales of weeks.  The offset of the source from
the dynamical center of the galaxy places an upper bound on its mass of
$10^{5} - 10^{6} \, M_{\odot}$, depending on its age, due to dynamical
friction \citep{kaaret01}.  These observations rule out a low
luminosity AGN or a very luminous supernova or hypernova remnant as the
source of the X-ray emission.  If the radiation from X41.4+60 is
isotropic and from an accreting object, then the mass of the accretor
must be in excess of $500 M_{\sun}$.  Combined with the mass upper
bound, this would place the object firmly in the ``intermediate-mass''
black hole category.

However, alternative explanations must be considered, particularly that
the ULXs radiate at super-Eddington luminosities \citet{begelman02} or
that the radiation is beamed, so the true luminosity is significantly
less than calculated assuming isotropic emission.  \citet{king01} have
suggested that ULXs are high-mass X-ray binaries in a short-lived phase
which occurs early in the life of almost every system and during which
mass is transferred from the companion to the compact object on the
thermal time scale of the companion -- leading to super-Eddington
accretion rates.  The X-ray emission in this phase is mechanically
beamed, producing high observed X-ray fluxes for observers near the
beaming axis.  However, the flux of X41.4+60 is so high that producing
it via mechanical beaming of a stellar-mass black hole is problematic. 
\citet{king01} specifically point to X41.4+60 as an example of a system
where their model may not apply.

\citet{mirabel99} pointed out that radio microquasars with their axes
aimed at us (``microblazars'') should occur in nearby galaxies. 
\citet{kording02} have suggested that the ULXs may be stellar-mass
black holes in which a relativistic jet produces most of the observed
X-ray flux.  Jets aligned nearly along our line of sight could then
produce high apparent X-ray fluxes via relativistic beaming.  If the
ULXs are microblazars, then they should be radio sources and the X-ray
and radio emission of the ULXs should be well correlated and highly
variable on time scales as short as days.  \citet{kaaret01} noted that
an unusual radio transient, 41.5+59.7 \citep{kronberg85}, lies within
the error box for X41.4+60.  This transient was detected in February
1981 at a 5 GHz flux of 7~mJy and not detected with an upper limit of
1.5~mJy in October 1983.  The sharp decrease in flux indicates that it
is not a supernova remnant.  Transient radio emission from an accreting
compact object would suggest the presence of a jet of relativistic
particles \citep{kaaret03}.

Recently, \citet{strohmayer03} found quasiperiodic oscillations (QPOs)
in the range 50--110~mHz which they identify as arising from X41.4+60. 
If the QPO frequency is limited by the orbital frequency around a
non-rotating black hole, then it would imply a mass limit of $2 \times
10^{4} \, M_{\odot}$.  The QPOs were discovered in XMM-Newton data and
confirmed in RXTE data.  The QPOs are only occasionally detected in the
RXTE data and there is no apparent correlation between QPO detections
and the source flux level.  Based on a comparison with the spectral and
timing properties of stellar-mass black hole X-ray binaries,
\citet{fiorito04} estimate a mass of the order of 1000$M_{\odot}$ for
the compact object producing the QPOs.  The QPOs were confirmed in a
longer XMM-Newton observation, but found at a somewhat higher frequency
of 113~mHz \citep{mucciarelli06,dewangan06}.

The position of X41.4+60 is within $1\arcsec$ of the position of the
infrared source and super star cluster MGG 11
\citep{kaaret04,portegies04a}.  Simulations of the dynamical evolution
of the cluster MGG 11 show that stellar collisions in its extremely
dense core may have led to numerous stellar collisions and the
formation of an intermediate mass black hole \citep{portegies04a}.

In order to obtain more information about X41.4+60, we monitored M82
for 240~days using the Proportional Counter Array (PCA) on the Rossi
X-Ray Timing Explorer (RXTE) with observations obtained typically every
2 days.  The PCA is not an imaging instrument, but is adequate to
measure the X-ray flux and rapid variability of the integrated output
of the entire galaxy.  We discovered an X-ray periodicity from M82 at
$62.0 \pm 2.5$ days which we have previously reported \citep{kaaret06}.
Additional details of the analysis are presented here.  We also found a
transient event where the X-ray flux of the galaxy increased by a
factor of three over the quiescent level over a period of a few days. 
After the onset of this remarkable X-ray flare, we obtained imaging
X-ray observations with the Chandra X-Ray Observatory and radio
observations with the Very Large Array (VLA).  We describe X-ray
observations with RXTE and Chandra in \S~2, the VLA data in \S~3,  and
discuss the results in \S~4.

\section{RXTE observations}

We obtained a total of 141 observations of M82 with a total effective
exposure of 281.2~ks using the Proportional Counter Array (PCA) on the
Rossi X-Ray Timing Explorer (RXTE) under program 90121 (PI Kaaret). Two
of the observations were made early to test our procedures for the
acquisition and analysis of realtime data.  The bulk of the
observations consisted of monitoring the galaxy typically every second
day during the period over which the VLA was in the A, BnA, or B
configuration during 2004 and 2005 and covered dates from MJD 53252 to
53490.

To enable target of opportunity observations with Chandra and the VLA,
we analyzed the real time data from RXTE soon after the data became
available on the RXTE ftp site.  The analysis was carried out with the
FTOOLS package (version 5.3.1) which is part of the HEAsoft software. 
After the data were retrieved, they were filtered to select good time
intervals such that Proportional Counter Unit (PCU) 2 was on, the
source was more than ten degrees above the horizon, the pointing offset
from the source was less than $0.01\arcdeg$, the satellite was out of
the South Atlantic Anomaly (SAA), and the electron contamination was
less than 0.1.  A background file was made for the same interval as the
observation with the model pca\_bkgd\_cmfaintl7\_eMv20031123.mdl and
the newest predicted SAA passage file downloaded at the same time as
the data.  The FTOOL saextrct was used to make both light curves and
spectra for the observation and the background, using only data from
PCU 2.  The spectra were fitted with a power-law model using the
spectral fitting program XSPEC, which is part of HEAsoft, to obtain a
measurement of the flux in the 2--10~keV band.  Our flux monitoring led
to the detection of an X-ray flare reaching a flux of $4.5 \times
10^{-11} \rm \, erg \, cm^{-2} \, s^{-1}$ in the 2--10~keV band on 26
Jan 2005 (MJD 53396.4).  On the basis of this flare detection, we
triggered target of opportunity observations with Chandra and the VLA.

\begin{figure}[tb]
\centerline{\includegraphics[width=3.0in]{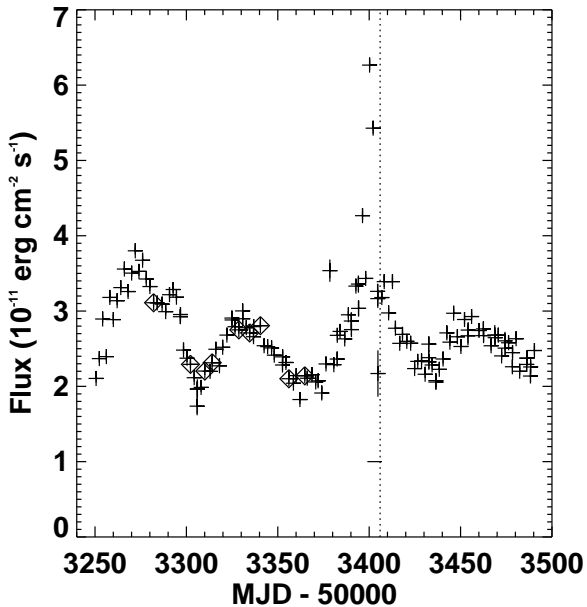}}
\caption{\label{xte_lc}  Light curves of M82 obtained the PCA on RXTE.
The plot shows the flux in the 2--10~keV band calculated for each
observation versus the observation date in MJD.  The diamonds indicate
the observations where QPOs were detected.  The vertical dashed line
indicates the time of the Chandra observation.  The solid horizontal
line indicates the interval over which VLA observations were obtained. 
A strong modulation in the X-ray flux near a period of 62 days is
apparent.  An X-ray flare occurred around MJD 53400.} \end{figure}

After the observing program was completed, we re-examined all of the
observations using the RXTE production data.  This had the advantages
that some data gaps in the real-time data were filled and, more
importantly, that the SAA passage history, rather than the SAA passage
predictions, could be used in estimating the PCA background.  To
produce an energy spectrum for each observation, we used the Rex script
\footnote{http://heasarc.gsfc.nasa.gov/docs/xte/recipes/rex.html}. We
produced spectra using only the top layer in PCU 2, including  channels
2-44 which cover the energy range 2-20~keV.  Response matrices were
made for each observation.  The expected background was estimated using
the same model mentioned above and a systematic uncertainty was
included in the background estimate
\footnote{http://lheawww.gsfc.nasa.gov/users/craigm/pca-bkg/bkg-users.html}.
The spectrum was fitted using XSPEC with a power-law model with
interstellar absorption with the column density fixed to $3 \times
10^{22} \rm \, cm^{-2}$.  The flux was calculated for the 2-10 keV
range. 

Fig.~\ref{xte_lc} shows the flux in the 2-10~keV band versus time. The
X-ray flare is obvious as the high flux points around MJD 53400.  The
peak flux occurred on MJD 53400.2 and reached a level of $6.3 \times
10^{-11} \rm \, erg \, cm^{-2} \, s^{-1}$ in the 2--10~keV band. The
duration of the flare is rather short, only a few days.  This is in
contrast to the light curve of M82 presented by \citet{rephaeli02},
where the duration of the high flux interval is at least 100 days.  
The photon index is generally between 2.2 and 2.5 and shows no strong
correlation with flux.

\begin{figure}[tb]
\centerline{\includegraphics[width=3.0in]{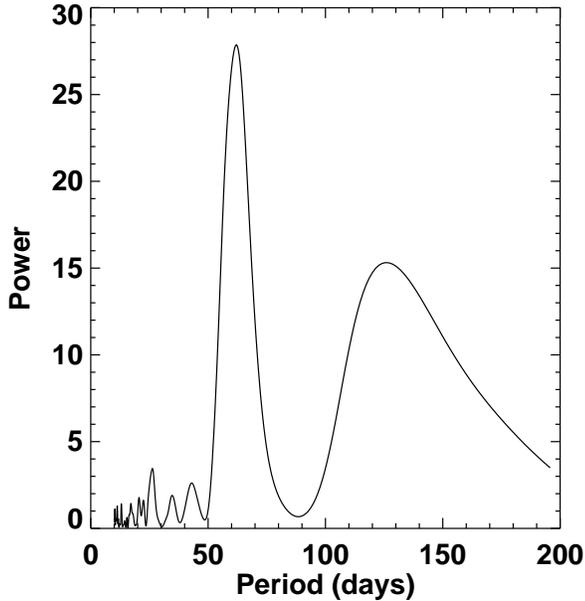}}
\caption{\label{periodogram}  Periodogram calculated from the flux
measurements of M82 obtained for individual observations.  A strong
peak is apparent at a period of $62.0 \pm 2.5$~days.  The periodogram
is calculated via the method of \citet{horne86} with the power
normalized by the total variance of the data.} \end{figure}

The light curve shows an apparent modulation with an amplitude of $1.6
\times 10^{-11} \rm \, erg \, cm^{-2} \, s^{-1}$ and a period near
60~days.  We calculated a periodogram, shown in Fig.~\ref{periodogram},
according to the method of \citet{horne86} with the power normalized by
the total variance of the data.  For the periodogram, we used our
entire RXTE data set.  There is a peak at a period 62.0~days with a
power of 27.9.  The probability of chance occurrence calculated
assuming a Poisson distribution, or white noise background, and taking
into account the number of trials is $1.1 \times 10^{-10}$,
corresponding to a  $6.7\sigma$ detection.  We estimate the 90\%
confidence error on the period to be 2.5~days.  There is a secondary
peak at 125.7~days with a power of 15.3.  Removing the two early
observations does not significantly affect the periodogram.  The peak
is then at $61.6 \pm 2.7$ days with a power of 27.5.  Removing the
observations during the X-ray flare near MJD 53400 does not shift the
peak, but increases its power.  With the points with fluxes above $4
\times 10^{-11} \rm \, erg \, cm^{-2} \, s^{-1}$ removed, the peak is
at $62.0 \pm 2.1$ days with a power of 32.8.  Removal of the two early
points and the high flux points leads to a peak is at $61.6 \pm 2.2$
days with a power of 32.8.

Accreting compact objects often exhibit red noise, or increasing timing
noise at decreasing frequency with a power-law dependence of power on
frequency.  Red noise can produce apparent periodicities at periods
comparable to the observation duration \citep{israel96}.  We tested the
significance of the observed signal using a red noise background. 
First, we fit a power-law to the power versus frequency relation for
periods shorter than 40 days \citep{vaughan05}.  We found a spectral
index of -0.78.  We then generated red noise with this spectral index
and with a mean and variance equal to the mean and variance of the
actual data.  This procedure is conservative because it includes the
variance due to the signal itself and also due to the X-ray flare.  We
generated simulated light curves using the {\tt rndpwrlc} routine of
the {\tt aitlib} IDL subroutine library\footnote{available at
http://astro.uni-tuebingen.de/software/aitlib} provided by the Institut
f\"ur Astronomie und Astrophysik of the Universit\"at T\"ubingen.  For
each light curve, we generated 2048 data points with uniform spacing
equal to one third of the average spacing of the main 139 observations
of our monitoring program and used only 141 points from the middle of
the simulated data with relative times matching the actual
observations.  Generating a simulated light curve longer than the
portion used helps eliminate the effects of red noise leakage.  The
simulated data were processed with the same procedures used to analyze
the real data.  We generated 100,000 trial light curves and searched
for cases where the power at a period of 80 days or less, corresponding
to three oscillation cycles in the main portion of the monitoring, was
greater than or equal to the observed value of 27.9.  We found only 2
cases, equivalent to a probability of chance occurrence of our observed
signal of $2\times 10^{-5}$.   


The Q value of the peak, the period of the peak divided by the full
width at half maximum power, in the periodogram is 4.5, which is fully
consistent with that expected for a periodic process given the
observation duration, but is only weakly constraining.  To attempt to
determine whether or not the signal is strictly periodic, we measured
the arrival times of the peak maxima.  We fit the data near each peak,
$\pm$20 days, with a second order polynomial.  We excluded the very
high flux points with fluxes above $4 \times 10^{-11} \rm \, erg \,
cm^{-2} \, s^{-1}$.  Comparing the peak times to an assumed 62 day
period, we find that the measured arrival times of first, second, and
fourth peaks are consistent with the predicted arrival time for
periodic signal within 2 days.  The third peak differs by 4 days. 
However, the polynomial fit is affected by two high flux points in the
decaying part of that peak (at MJD 53408.2 and 53412.7).  If we fit the
data within $\pm$11 days of that peak, we find that the peak arrival
time agrees within 1 day of the time predicted for a periodic signal. 
We note that we had observations only every second day and there are
variations in the flux from other sources in M82 which are not related
to the 62 day periodicity and which lead to uncertainty in the time of
arrival determination.  To estimate the uncertainty in the polynomial
fits, we varied the width of the data windows and the placement of the
data windows, i.e.\ offsets the windows with respect to the peak.  We
conclude that the peak arrival times are accurate to about 2 days. 
Therefore, the arrival times of the peaks are consistent with being
periodic at an interval of 62.0 days.  However, additional observations
are essential to test the periodic nature of the signal.

There is archival RXTE data from a previous set of observations of M82
made in 1997.  However, the time coverage of that data set is uneven
and generally less dense than in our monitoring program.  There is a
hint of one peak in the 1997 light curve around MJD 50585 which has a
shape similar to the four seen in our data.  However, the 1997 data
alone is insufficient to confirm or deny the periodicity.  If we make a
periodogram combining the two data sets, we find a peak at 61.5 days.

\begin{figure}[tb]
\centerline{\includegraphics[height=4.5in]{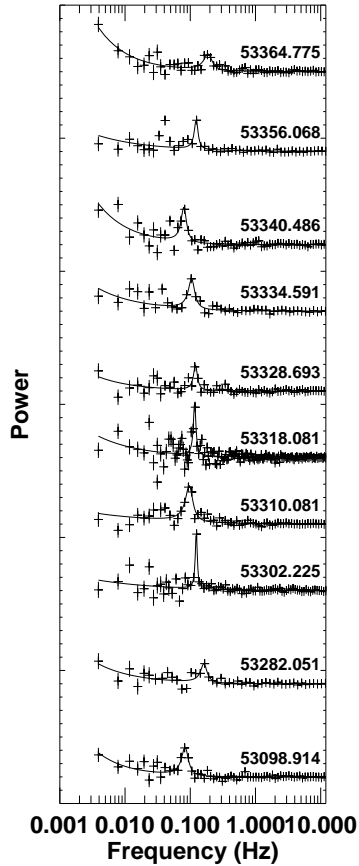}}
\caption{\label{xte_powspec}  Selected power spectra of M82 obtained
with the PCA on RXTE.  Each power spectrum is labeled with the MJD of
the observation date.  The power spectra shown are those with QPOs
detected with a single trial significance of $4\sigma$ or greater.  The
QPO frequencies lie in the range from 80 to 190~mHz.} \end{figure}

To examine the rapid variability of the source, 152 event files were
examined individually according to the following procedure using a
program written in the Interactive Data Language (IDL).  Events in the
2-10~keV energy band were selected and split into segments of 256~s. 
An FFT with a time resolution of $2^{-6}$~s was calculated for each
segment.  The FFTs were added incoherently.  The resulting total power
spectrum was logarithmically rebinned with the bin width adjusted on a
case by case basis.  The IDL routine lmfit was then used to fit a model
consisting of an exponential background and, if necessary, a Lorentzian
peak to model a QPO.

Of 152 event files, 10 contained QPOs with a single trial significance
of $4\sigma$ or greater. The power spectra from observations with QPO
detections are plotted in Fig.~\ref{xte_powspec}.  The centroid
frequencies of the QPOs range from 80 to 190~mHz.  The maximum
modulated flux in the QPO is $1.9 \times 10^{-12} \rm \, erg \, cm^{-2}
\, s^{-1}$ in the 2--10~keV band.

\subsection{Chandra observations}

After detection of the large X-ray flare in the PCA monitoring, a
Chandra Target of Opportunity Observation was triggered.   The request
was sent to the Chandra X-ray Center on 26 Jan 2005 and the observation
(ObsID 6097; PI Kaaret) began on 4 Feb 2005 at 23:34:40 UT (MJD
53405.982).  The observation was made using the Advanced CCD Imaging
Spectrometer spectroscopy array (ACIS-S; Bautz et al.\ 1998) and the
High Resolution Mirror Assembly (HRMA).   We obtained 52,773~s of
useful exposure.   The ACIS-S was used in imaging mode.  The high flux
anticipated from the source would have caused severe pile-up if the
source were placed on-axis and the standard observing mode were used.
To reduce the effect of pile-up, the target was offset $3\arcmin.57$
along the Y-detector coordinate.  This placed the target off axis where
the point spread function broadens and the X-ray flux is spread over
multiple pixels in the CCD detector which reduces pile-up.  In
addition, we operated only the S3 chip and employed a 1/8 sub-array
mode to reduce the frame time to 0.441~s and further reduce the
pile-up.  To check if pile-up was significant in the data acquired, we
calculated the count rate in the $3\times 3$ pixel cell at the peak of
the brightest source in the image.  The rate was 0.18 counts/frame,
giving a pile-up fraction of 3.2\% within this $3\times 3$ pixel cell. 
The counts in the peak $3\times 3$ pixel cell are 70\% of the total
source counts, so the overall pile-up fraction is 2.2\%.  Therefore,
the effect of pile-up on the timing and power spectra is small.

\begin{figure*}[tb]
\centerline{\includegraphics[width=5in]{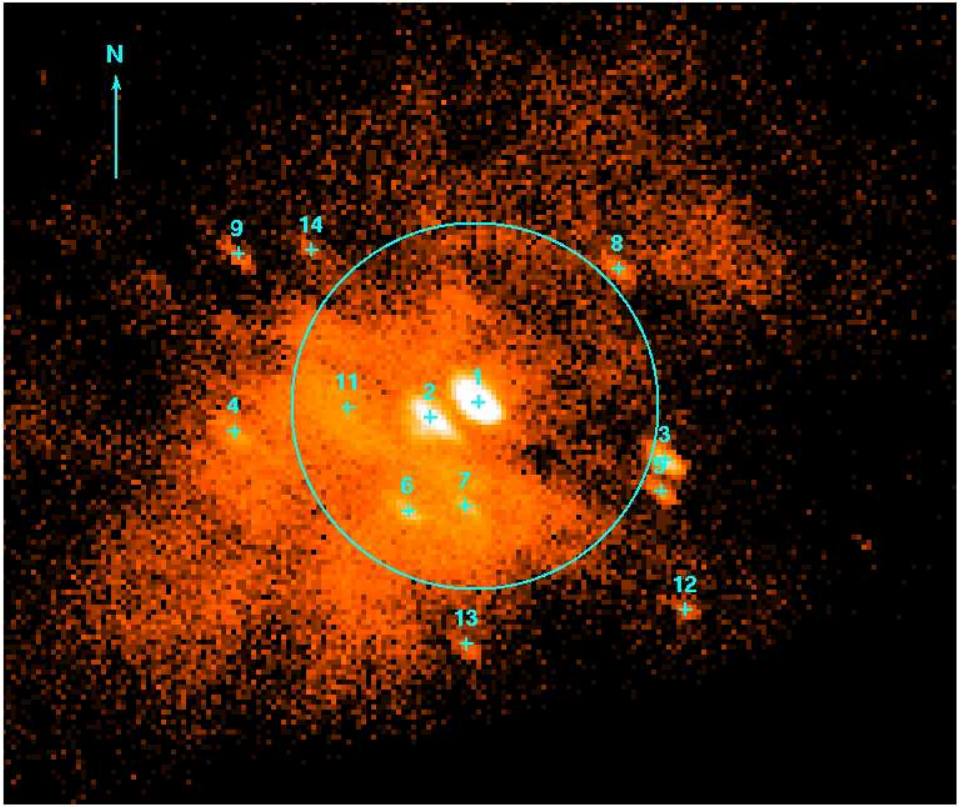}}
\caption{\label{chandra_image}  X-ray image of M82. The image includes
all events in the 0.3--8~keV band.  The light blue crosses indicate the
positions of point sources.  The brightest source is X41.4+60 and is
located at the center of the image and labeled source \# 1.  The second
brightest source is X42.3+59 is labeled source \# 2.  The light blue
circle indicates the extraction region used for the analysis of
XMM-Newton data by \citet{strohmayer03}.   Note that both of the
brightest sources and several additional weaker sources lie within that
extraction region.  The arrow points north and has a length of
$10\arcsec$.} \end{figure*}

The Chandra data were subjected to standard data processing and event
screening (ASCDS version 7.5.0.1 using CALDB version 3.0.0).  The total
rate on the S3 chip was between 2.5--3.8~c/s for the entire observation
and did not show any strong background flares.  We constructed images
for the 0.3--8~keV and 2--8~keV bands using all valid events and used
the {\it wavdetect} tool in the {\it CIAO} version 3.1 Chandrda data
analysis package to search for X-ray sources.  We found that use of the
higher energy band seemed to produce more reliable point source
detections and that the diffuse emission had less influence on the
calculated source positions.  Therefore, we used the source positions
found from the 2--8~keV data in the remainder of the analysis. 

Fig.~\ref{chandra_image} shows the image in the 0.3--8~keV band. 
Sources with detection significance of $7\sigma$ or higher are marked. 
The point sources appear as ellipses due to the off-axis pointing of
the observation.  The circle in the figure represents the extraction
region used by \citet{strohmayer03} in their analysis of the XMM-Newton
data.  Several bright X-ray sources fall in or near the extraction
region and must have contributed to the energy and timing spectra
attributed to X41.4+60.
  
\begin{figure}[tb]
\centerline{\includegraphics[width=3.0in]{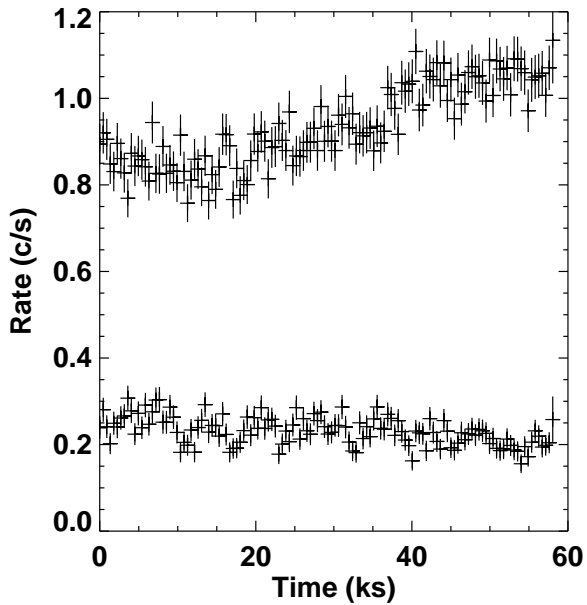}}
\caption{\label{chandra_lc}  Light curves of the two brightest sources
detected in the Chandra data.  The upper curve is for the brightest
source, X41.4+60.  It shows a gradual increase on time scales of tens
of kiloseconds.  The lower light curve is for the second brightest
source, X42.3+59.  It shows variability on time scales of kiloseconds. 
The light curves include events in the 0.3--10~keV band.  The time bins
are 450~s.} \end{figure}

\begin{figure}[tb] \centerline{\includegraphics[width=3.0in]{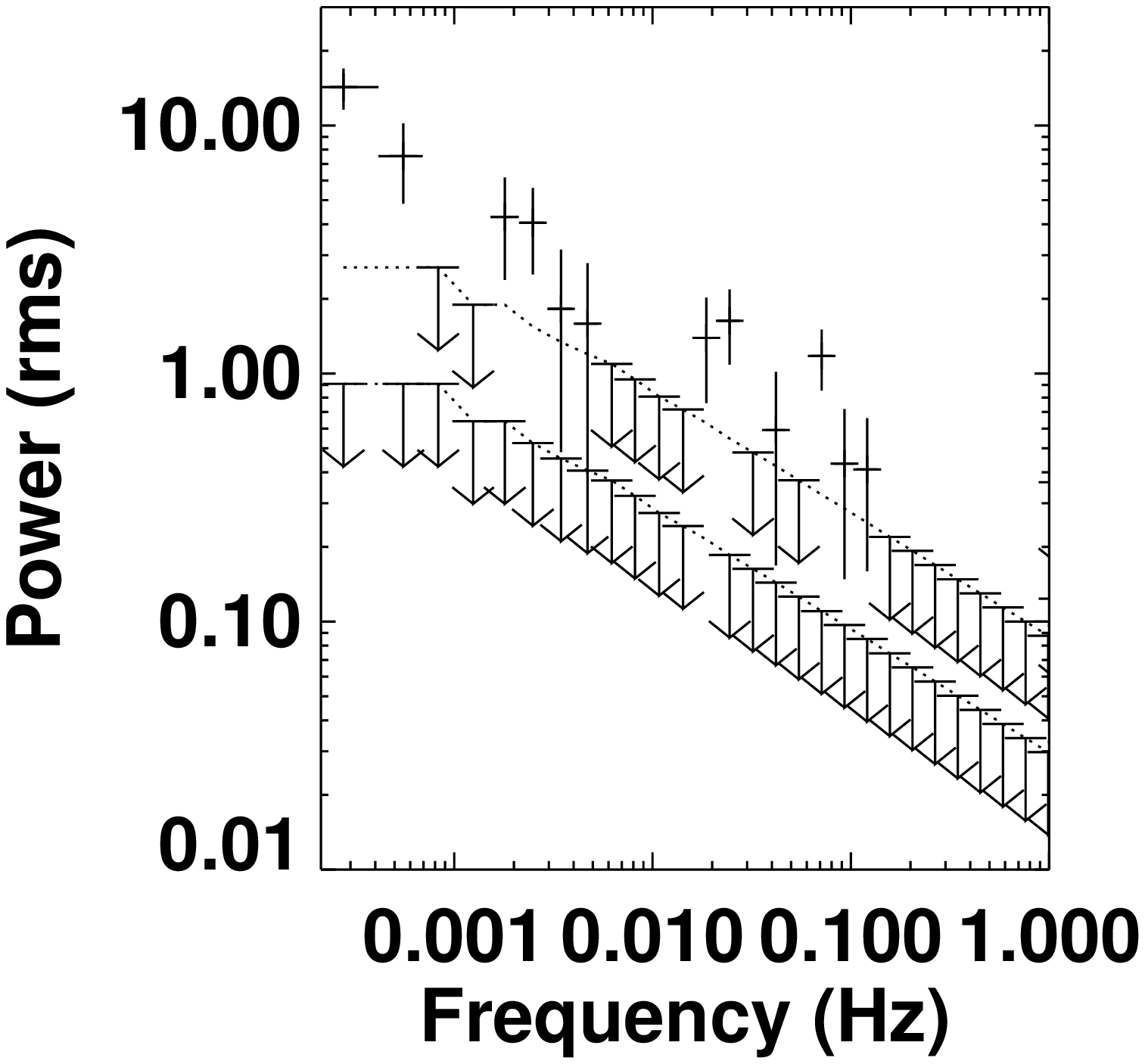}}
\caption{\label{chandra_powspec}  Power spectra of the two brightest
sources detected in the Chandra data.  The lower curve is for the
brightest source, X41.4+60.  Its power spectrum is consistent with
Poisson noise over the frequency range plotted.  The upper curve is for
the second brightest source, X42.3+59.  It shows significant timing
noise at low frequencies and a possible QPO near 70~mHz.  The power
spectra are calculated from events in the 2--10~keV band.} \end{figure}

The two brightest sources have fluxes which are an order of magnitude
greater than any other source in the field.  They are of particular
interest and are the topic of the reminder of our Chandra analysis. 
The brightest source is CXOU J095550.2+694047, the known ULX in M82
which was discussed in the introduction.  Following \citet{kaaret01},
we will refer to this source as X41.4+60.  The second source is located
at  of $\alpha=$ 09h 55m 51s.040 and $\delta=$ +69$\arcdeg$ 40$\arcmin$
45$\arcsec$.49 (J2000).  This source is likely the same as CXOM82
095551.4+694045 in \citet{matsumoto01} and may be the same as CXOM82
095551.4+694043 in \citet{griffiths00}.  Following the convention of
naming sources in M82 by their offset from $\alpha=$ 09h 55m 00s,
$\delta=$ +69$\arcdeg$ 40$\arcmin$ 00$\arcsec$ (B1950), we refer to
this source as X42.3+59.  We note that \citet{strohmayer03} refer to
the brightest source in M82 as M82 X-1.  They assumed that this source
is the same as X41.4+60 and  ascribed it to be the origin of the QPOs
found in the XMM-Newton and RXTE data.  However, as discussed further
below, X42.3+59 lies within the extraction radius used by
\citet{strohmayer03} to define the X-ray counts from M82 X-1. 
Therefore, their source M82 X-1 may include contributions from both 
X41.4+60 and X42.3+59.

We extracted photon lists and created light curves and energy spectra
for both sources using standard CIAO tools.  The light curves are shown
in Fig.~\ref{chandra_lc}.  The light curves represent the count rate in
the 0.3--10~keV band.  The upper curve is for the brighter source,
X41.4+60.  The source appears to vary gradually on time scales of
several thousand seconds.  The flux of X42.3+59 appears to have
variations on faster times scales, but is relatively constant on long
time scales.

Fig~\ref{chandra_powspec} shows the timing power spectra for both
sources derived from counts in the 2--10~keV band.  The plotted
spectrum for each source is the sum of 16 individual power spectra with
8192 points each.  The summed power spectrum was rebinned into
logarithmically spaced bins before plotting.  The sampling time was
equal to the ACIS frame time of 0.441~s.  In the power spectra plot,
the curve for X41.4+60 is the lower one.  The timing noise from
X41.4+60 is consistent with Poisson fluctuations at frequencies from
0.001~Hz to 1~Hz.  In contrast, X42.3+59 shows significant noise near
1~mHz and a possible QPO near 70~mHz.  Taking into account the number
of trials, i.e.\ the number of frequency bins in the power spectrum,
the QPO at 70~mHz has a chance probability of occurrence of $9\times
10^{-3}$.

\begin{figure}[tb]
\centerline{\includegraphics[width=3.0in]{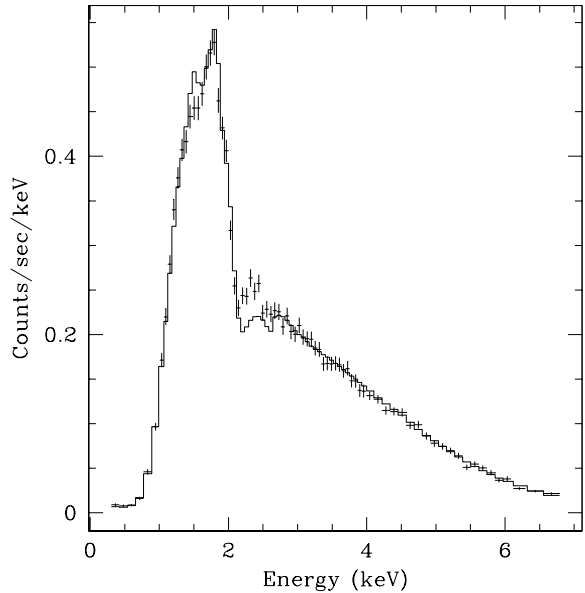}}
\caption{\label{chandra_spec1}  Energy spectrum of X41.4+60.  The
spectrum is adequately fitted by a power-law with interstellar
absorption except for a region just above the Ir M edge at 2.1~keV
where the response matrix at large off-axis angles is not well
determined.} \end{figure}

The energy spectrum for X41.4+60 is shown in Fig.~\ref{chandra_spec1}.
We fitted the spectrum using the Sherpa fitting package which is part
of CIAO.  The spectrum is adequately fitted by a power-law with
interstellar absorption except for a region where there are significant
residuals just above 2.1~keV.  There is a strong feature in the
instrument response at 2.1~keV corresponding to the Ir M edge.  An
analysis of grating data presented by Herman Marshall at the 2003
Chandra Calibration
Workshop\footnote{http://cxc.harvard.edu/cal/Hrma/on\_axis\_effarea.html}
shows that a $\sim 10\% 
$ correction is needed in the HRMA response at
energies just above the Ir-M edge.  Since this correction may depend on
the off-axis angle of the source and does not appear to be adequately
modeled in the current response matrices, we have chosen to remove the
data for energies from 2.1--2.7~keV in performing the fits.

For X41.4+60,  we find that the energy spectrum is adequately fitted
with  a power-law with interstellar absorption giving $\chi^2$/DoF =
72.8/66.  The best fit parameters were a photon index of $\Gamma = 1.67
\pm 0.02$ and an equivalent hydrogen absorption column density of
$N_{H} = 1.12 \pm 0.02 \times 10^{22} \rm \, cm^{-2}$.   The source
flux was  $1.3 \times 10^{-11} \rm \, erg \, cm^{-2} \, s^{-1}$ in the
0.3--7~keV band and  $1.5 \times 10^{-11} \rm \, erg \, cm^{-2} \,
s^{-1}$ in the 2--10~keV band.  We note that an absorbed multicolor
disk blackbody model is strongly rejected with $\chi^2$/DoF = 4.0. 
Addition of a multicolor disk blackbody to the power-law model does not
significantly improve the fit.   For disk temperatures greater than
0.5~keV, the normalization of the disk blackbody component is $N < 1$. 
However, due to the significant absorption at low energies, we cannot
rule out the possible presence of a cool disk component with a
temperature below 0.3~keV.  Addition of an Fe-K emission line also does
not improve the fit.  Taking a gaussian line profile with a centroid of
6.55~keV and a width of 0.33~keV to match the line reported by
\citet{strohmayer03}, we place a 99\% confidence level upper bound of
70~eV on the equivalent width of the line.  This is below the 
equivalent widths found from the XMM-Newton data.

For X42.3+59, the source flux was  $4.4 \times 10^{-12} \rm \, erg \,
cm^{-2} \, s^{-1}$ in the 0.3--7~keV band and  $6.9 \times 10^{-12} \rm
\, erg \, cm^{-2} \, s^{-1}$ in the 2--10~keV band. A single power-law
with interstellar absorption did not give an adequate fit.  We found an
adequate fit with the sum of two power-laws with independent
interstellar absorption, with $\Gamma_{1} = 1.33 \pm 0.07$,
$(N_{H})_{1} = 2.8 \pm 0.2 \times 10^{22} \rm \, cm^{-2}$, $\Gamma_{2}
= 4.13 \pm 0.01$, $(N_{H})_{2} = 9.0 \pm 0.9 \times 10^{21} \rm \,
cm^{-2}$, and $\chi^2$/DoF = 156.0/162.  We also found an adequate fit
with the absorbed sum of a multicolor disk blackbody plus a power-law. 
The best fit parameters were $N_{H} = 2.29 \pm 0.11 \times 10^{22} \rm
\, cm^{-2}$, $\Gamma = 1.19 \pm 0.06$, disk inner temperature $kT_{in}
= 0.072 \pm 0.002 \rm \, keV$, disk normalization $N = 3.2 \times
10^{7}$, and $\chi^2$/DoF = 162.1/163.  This disk normalization would
imply a black hole mass in excess of $4 \times 10^{5} M_{\sun}$.

\begin{figure*}[tb]
\centerline{\includegraphics[scale=0.6,angle=0]{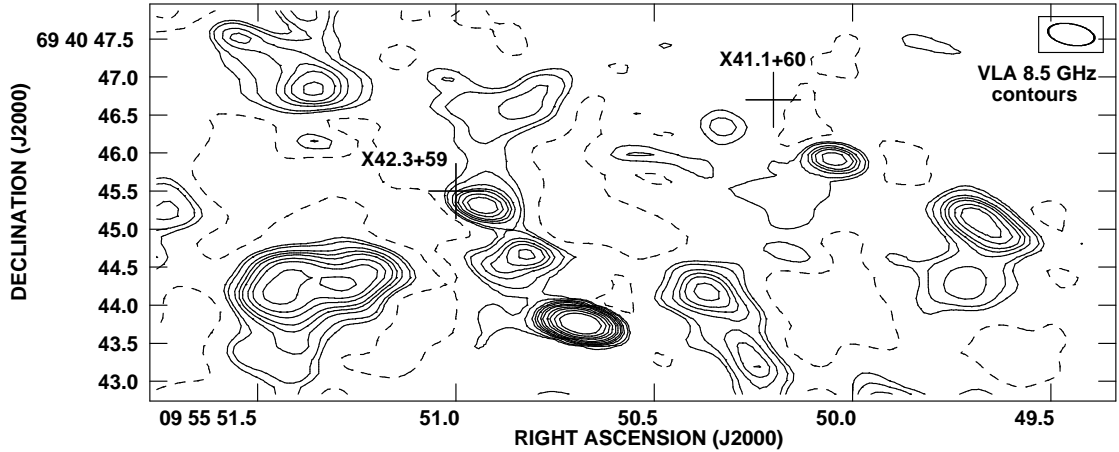}}
\caption{\label{radio_m82} VLA 8.5 GHz image of a central region in M82
containing the two X-ray brightest sources (indicated by crosses of
length 0.7\arcsec, corresponding to Chandra positional errors). There
is a possible (but low signal-to-noise) radio source at a displacement
of 1.2\arcsec~from the brightest X-ray source, X41.4+60.  The flux of
this radio source is less than 1 mJy.  The second brightest X-ray
source, X42.3+59, has a position which is coincident with the radio
source 42.21+59.0 (its flux is $\sim$2.5 mJy and detected here with a
signal-to-noise of 30).  Contours show radio emission at levels of -3,
3, 5, 9, 12, 15, 20, 25, 35, 45, 60 and 80 times the rms level of 0.08
mJy/beam.  The resolution of this image is 0.62\arcsec~$\times$
0.28\arcsec, PA=79.8\arcdeg.} \end{figure*}

\section{VLA Observations}

Four separate radio observations using the Very Large Array (VLA) of
the  National Radio Astronomy Observatory (NRAO)\footnote{The NRAO is a
facility of the National Science Foundation operated under cooperative
agreement by Associated Universities, Inc.} were made of M82 after
detection of X-ray flare.  The first observation occurred on 29 Jan
2005 (MJD 53399) and the last on 5 Feb 2005 (MJD 53406).  The radio
observations began one day before the peak X-ray flux and continued to
6 days after.  All observations were made at 8.5~GHz, with integration
times ranging from 1 to 2.5 hours and the array was in its hybrid BnA
configuration. J1331+305 was used as a flux calibrator, and J0958+655
was observed every 15 minutes to calibrate the interferometer phases.
Standard procedures for calibration, editing and imaging were carried
out using the Astronomy Image Processing Software (AIPS) of NRAO. 

Images of M82 were produced for each  observation day as well as a
composite image produced from the combination of all data, shown in
Fig.~\ref{radio_m82}.  In order to study the compact sources in M82 at
8.5 GHz, the extended emission larger  than 4\arcsec~was filtered out
by limiting the corresponding  (u,v) data range to baselines longer
than 50 k$\lambda$. The resulting  resolutions of all images are
subarcsecond (approximately $0.6\arcsec \times 0.3\arcsec$).  

We detected a low signal-to-noise radio source in two of the four
observations at a position which is $1.2\arcsec$ from the position
derived for X41.4+60 in our TOO observation, but only $0.7\arcsec$ from
the position quoted for X41.4+60 in \citet{kaaret01}.  Given the
astrometric uncertainty of Chandra, we conclude that the source is a
possible radio counterpart to X41.4+60.  This radio source is
unresolved in all observations and has a position of (J2000): 09 55
50.33, 69 40 46.3. The 8.5 GHz flux density of this source is $0.9 \pm
0.1$~mJy on 1 Feb 2005 and $0.6 \pm 0.1$~mJy on 3 Feb 2005.  The source
is not detected on 29 Jan 2005 with an upper limit of $0.3 \pm 0.1$~mJy
and a marginal detection of $0.5 \pm 0.1$~mJy on 5 Feb 2005.
\citet{rvzga04} report the only other detection of a radio source at
this position, 41.62+59.9 (J1950 coordinates name), with a flux density
of $1.6 \pm 0.1$~mJy at 8.5~GHz.  The source is $0.8\arcsec$ from the
radio transient reported by \citet{kronberg85}.

Radio emission is also detected from a source which falls within the
$0.7\arcsec$~error radius of X42.3+59.  This radio source, known as
42.21+59.0, has been detected previously by a number of radio studies
of compact sources in M82.  We measure the flux density of 42.21+59.0
to be $2.9\ pm 0.1$~mJy during the first three observations and $2.4
\pm 0.1$~mJy on 5 Feb 2005.  Values for the 8.5~GHz flux density of
this source range from 1.9~mJy \citep{huang94} to 6.7~mJy
\citep{rvzga04}.  The range of values is most likely due to the
differing interferometric arrays used for observation (which are
sensitive to different spatial scales of radio emission) and the
different techniques used to account for the extended background
emission surrounding the source. Based on its flat or slightly positive
spectrum (between 5 and 15~GHz), extended size, the presence of OH and
water masers and radio recombination line emission, 42.21+59.0 is
believed to be a giant HII region containing as many as 94 O5-type
stars \citep{ak98,mm02,rvzga04}.

\section{Discussion}

\subsection{X-Ray Flare}

The fluxes measured by RXTE for the observations before and after the
Chandra observation are consistent with $3.2 \times 10^{-11} \rm \, erg
\, cm^{-2} \, s^{-1}$ in the 2--10~keV band.  The total flux in the
entire Chandra image is $3.0 \times 10^{-11} \rm \, erg \, cm^{-2} \,
s^{-1}$ in the 2--10~keV band, which is consistent within the errors on
the flux estimate and the relative calibration uncertainty between RXTE
and Chandra.  Of the total Chandra flux in this band, 50\% arises from
X41.4+60 and 23\% arises from X42.3+59.  The peak flux measured by RXTE
during the flare was $6.3 \times 10^{-11} \rm \, erg \, cm^{-2} \,
s^{-1}$ on MJD 53400.2.  Assuming that a flux of $1.5 \times 10^{-11}
\rm \, erg \, cm^{-2} \, s^{-1}$ arises from the remainder of the
galaxy, the flux of the point source producing the flare would be $4.8
\times 10^{-11} \rm \, erg \, cm^{-2} \, s^{-1}$, equivalent to a
isotropic luminosity of $7.6 \times 10^{40} \rm \, erg \, s^{-1}$ at a
distance of 3.63~Mpc.  This is consistent with the highest fluxes
measured with Chandra from X41.4+60 \citep{kaaret01}. It appears likely
that X41.4+60 produced the flare observed with RXTE.

The energy spectrum that we find for X41.4+60 is best described by a
single absorbed power-law.  We find no evidence for thermal emission
from a disk or for Fe-K line emission.  The spectrum, a featureless
power-law with a relatively hard photon index, is reminiscent of the
spectra of blazars, but the low level of radio emission, even during
the X-ray flare, rules out a blazar identification for the source. 
Adopting the maximum flux that we observe for the radio source
41.62+59.9 as an upper limit on the radio flux of X41.4+60, we find
that the X-ray to radio flux ratio is more than an order of magnitude
larger than found for any known blazar \citep{landt01}.  The absence of
strong radio emission during the flare argues against it being a
relativistically beamed jet source \citep{kording02}.

It has been suggested that ULXs may be mechanically beamed
\citep{king01} or radiating at super-Eddington rates
\citep{begelman02}.  The flux from X41.4+60 in the 2-10 keV band at the
peak of from flare corresponds to a luminosity of $7.6 \times 10^{40}
\rm \, erg \, s^{-1}$ at a distance of 3.63~Mpc, equivalent to the
Eddington limit for a 580~$M_{\odot}$ object.   We note that this is
the observed (absorbed) flux.  Using the best fit spectral model from
the Chandra observation, this 2-10 keV flux corresponds to an
unabsorbed flux of $8.7 \times 10^{-11} \rm \, erg \, cm^{-2} \,
s^{-1}$ in the 0.3-10 keV band for a luminosity of $1.4 \times 10^{41}
\rm \, erg \, s^{-1}$ or $1.3 \times 10^{-10} \rm \, erg \, cm^{-2} \,
s^{-1}$ in the 0.1-20 keV band for a luminosity of $2.1 \times 10^{41}
\rm \, erg \, s^{-1}$.  The latter luminosities corresponds to the
Eddington limits for 1060~$M_{\odot}$ and 1600~$M_{\odot}$ objects,
respectively.  Given that the masses of known stellar-mass black holes
are all below 20~$M_{\odot}$ \citep{mcclintock03}, this would require
the inferred luminosity to exceed Eddington by a factor of least 30 and
more likely 50-80.  However, models of `funnel' shaped disks, as
required for mechanical beaming, indicate that the maximum flux
enhancement is only a factor of $\approx 5$ \citep{misra03}.  This
falls far short of the enhancement needed to explain X41.4+60 as a
standard black hole.  The formation of photon bubbles in magnetized,
radiation pressure dominated disks may also lead to super-Eddington
luminosities.  Originally, the potential to reach luminosities up to a
factor of 100 in excess of Eddington were suggested
\citep{begelman02}.  However, the best currently available calculations
indicate enhancement factors of a few \citep{ohsuga05,watarai05} or
less than 10 \citep{ruszkowski03}.

The question of whether or not black holes can radiate at substantially
super-Eddington luminosities can also be addressed by examination of
Galactic X-Ray binaries and active galactic nuclei (AGN). 
\citet{collin04} find that the bolometric luminosities of AGN saturate
at a few times the Eddington luminosity.  In a sample of 578 type 1
AGN, \citet{warner04} find that only the most extreme objects have
Eddington ratios approaching 10.  We note that errors in black hole
mass measurement and the AGN distances may contribute to the highest
Eddington ratios quoted.  For Galactic X-ray binaries with measured
black hole masses, the highest luminosities observed are only a few
times Eddington \citep{mcclintock03}.  It seems that the maximum
luminosities produced by accretion disks in nature exceed the Eddington
limit generally by no more than a factor a few and in the most extreme
objects by less than a factor of 10.

Given these considerations, it appears unlikely that X41.4+60 can be
interpreted as a standard stellar-mass black hole.  It's extremely high
luminosity and strong variability identify it as, perhaps, the best
intermediate-mass black hole candidate amongst the ULXs.

\subsection{X42.3+59}

Our Chandra observation found X42.3+59 at the highest flux yet seen. 
The 0.3-10~keV flux corresponds to an isotropic luminosity of $1.1
\times 10^{40} \rm \, erg \, s^{-1}$, a dramatic increase in brightness
over what was seen in previous observations.  The total 2-10~keV flux
found from the central region of M82 during the XMM-Newton observation
where  \citet{strohmayer03} detected a QPO was $2.1 \times 10^{-11} \rm
\, erg \, cm^{-2} \, s^{-1}$.  It is possible that X42.3+59 contributed
one third of the total XMM-Newton flux and was  the origin of the QPOs.
The required modulation would be 30\%, which is comparable to rms
fractions measured for low frequency QPOs, in the range 0.6-2~Hz, from
GRS 1915+105 \citep{morgan97}.  We note that our Chandra timing power
spectrum of X42.3+59 shows significant variability near 1~mHz and a
possible QPO at 70~mHz, similar to that seen with XMM-Newton, while the
timing noise from X41.4+60 is consistent with Poisson fluctuations in
this frequency range.  This may suggest that X42.3+59 and not X41.4+60
is the source of the QPOs detected with XMM-Newton and RXTE.

If the radiation is isotropic, then the minimum black hole mass for 
X42.3+59 is 85~$M_{\odot}$.  Black hole X-ray binaries tend to produce
QPOs while in the low/hard state, which corresponds to luminosities
below 10\% of the Eddington limit.  If X42.3+59 is the source of the
QPO detected from the central region of M82 and was is at a similar
luminosity while the QPO is being produced, then the estimated black
hole mass is of order 1000~$M_{\odot}$.  This may indicate the presence
of a second intermediate-mass black hole candidate in M82.

\subsection{X-Ray Periodicity}

We have found an apparent 62 day periodicity in the X-ray light curve
of M82, which likely arises from X41.4+60, the brightest X-ray source
in the galaxy.  This periodicity could be due to orbital modulation,
superorbital modulation, or an aperiodic phenomena.

The X-ray emission of several black hole X-ray binaries is modulated at
the orbital period.  In particular, Cygnus X-3 presents a clear case,
known since 1980 \citep{elsner80}, of a modulation in X-rays at the
orbital period.  More recently, \citet{boyd01} found a modulation at
the 1.7 day orbital period of LMC X-3, a black hole X-ray binary with a
high mass companion.  Also, recent results on systems with giant
companion stars suggest that such modulations are more common than in
low-mass systems.  \citet{smith02} found modulations at 12 days for 1E
1740.7-2942 for 18 days for GRS 1758-258 which they interpret as
orbital (they also find superorbital modulations at around 600 days),
and \citet{corbet03} found a 24.06 day period in X-rays from GX 13+1,
which he interprets as an orbital modulation.  The modulations from
these systems are roughly sinusoidal, but tend to have sharper peaks
and troughs than pure sinusoids.  The range of shapes is consistent
with the shape of the modulation that we observe from M82.  X-ray
modulation at the orbital period appears common for systems with giant
companions. 

There are so-called super orbital modulations in the X-ray emission
from X-ray binaries.  For neutron star systems, the super orbital
periods include the range around 62 days.  However, to produce the flux
observed from X41.4+60 from an accreting neutron star would required a
beaming factor of larger than 300.  This is much larger than has been
suggested to be theoretically possible in mechanically beamed models,
see above, and would require a beam with an angular spread of less than
$0.2\arcdeg$.  For black hole X-ray binaries, the shortest superorbital
period is 162 days (from SS 433 for which there is no dynamical mass
measurement confirming its black hole status) and the superorbital
periods are more typically in the 300-600 day range (for Cyg X-1, 1E
1740.7-2942, and GRS 1758-258). Therefore, the observed 62 day period
lies outside the range of observed superorbital periods for black hole
X-ray binaries. We cannot exclude the possibility that the process
producing the 62 day modulation could be aperiodic.  However, we are
not aware of any process which would produce aperiodic, but relatively
coherent, behavior with the time scale and modulation amplitude seen.
We conclude that the periodicity in the RXTE light curve likely
indicates the orbital period of X41.4+60.   

Possible orbital periodicities have been reported for a few other
ULXs.  The previous best candidate is CG X-1 in the Circinus galaxy
\citep{bauer01} for which (almost) three periods of 7.5~hours were
detected in an observation of 16.7~hours. The period was confirmed in
additional (shorter) Chandra observations and a BeppoSAX observation of
14.4~hours \citep{bianchi02}, and also in an XMM-Newton observation of
28.9~hours \citep{weisskopf04}.  The period is consistent with a main
sequence companion with a mass near 1~$M_{\odot}$ or a similarly low
mass core He-burning companion.  The light curve of CG X-1 resembles
that of an AM Herculis type cataclysmic variable star, with sharp
transitions between high and low flux phases, and the source lies
relatively close to the plane of the Milky Way $b = -3.8\arcdeg$, so it
may be a foreground AM Her system \citet{weisskopf04}.

Other reported detections of orbital modulations from ULXs are of
substantially lower significance. \citet{sugiho01} report possible
31~hr or 41~hr periods from a 154~hr ASCA observation of IC 342 X-1,
but the signal is relatively weak and even extraction of the best
period is ambiguous.  \citet{liu02} report a possible 2.1~hr period
from M51 X-7, but only two cycles were detected in a single Chandra
observation of 4.2~hr.  \citet{dewangan05} report a periodicity of
1.6~hr from the same M51 source in a 5.8~hr XMM-Newton observation
obtained 2.5 years after the Chandra observation, so it appears
unlikely that the modulation is related to the binary orbit and,
instead, is either a quasiperiodic or red-noise modulation of the flux.
\citet{david05} report a suggestion of a period of 8-10~hr from a ULX
in the elliptical galaxy NGC 3379, but this is based on a 9~hr
observation -- a single orbital period.  Quasiperiods have been seen
from several ULXs including a $\sim 2$~hr quasiperiod from a source in
NGC 7714 \citep{soria04} and a $\sim 1-2$~hr quasiperiod from a source
in M74 \citep{krauss05}.  However, these quasiperiods are not stable
enough to be interpreted as orbital modulations.

We note that very few ULXs have been monitored in a manner which would
permit detection of modulations on time scales of tens of days. 
Indeed, the program described here is the only such ULX monitoring of
which we are aware.  Therefore, the lack of detections of modulations
on time scales of tens of days may simply be an observational artifact.
Monitoring programs extending over 100-300 days with observations every
day or few days would be of great interest.  We note that the best
available optical constraint on the nature of a ULX companion star
comes from an HST spectrum from which \citet{liu04} identify the
companion to NGC 5204 X-1 as a B0 Ib supergiant.  They predict an
orbital period on the order of 10 days.  This provides additional
motivation for searching for X-ray modulations from ULXs with time
scales of tens of days.

\subsection{Nature of the ULX companion star and formation of the ULX}

We note that X41.4+60 is likely a Roche-lobe overflow system because
wind-fed accretion cannot provide a sufficiently high mass accretion
rate \citep{kaaret04}.  In this case, the orbital period directly
constrains the average density of the companion star according to the
equation $\rho \simeq 115 (P/{\rm hours})^{-2} \rm \, g \, cm^{-3}$
\citep{frankbook}.  If the orbital period is 62 days, then the mean
density of the companion star is $5 \times 10^{-5} \rm \, g \,
cm^{-3}$.  This excludes main sequence stars, but would be compatible
with giant stars or supergiant stars.  If, instead, the 125.7~day
periodicity is the orbital period, then the companion star density
would be $1.3 \times 10^{-5} \rm \, g \, cm^{-3}$ and the companion
star must, again, be a giant or supergiant star.

As noted above, the only other ULX with a reasonably firm
identification of the spectral type of the companion is NGC 5204 X-1
for which \citep{liu04} find a spectral type of B0 Ib based on a
far-ultraviolet spectrum.  This may suggest that giant or supergiant
companion stars are common in ULXs.  However, more companion spectral
types must be determined to drawn any firm conclusions.  The spectral
types for the companions to some other ULXs are been estimated using
optical colors, but these classifications are suspect because optical
emission from the accretion disk could affect the colors
\citep{kaaret05}.  The density of $5 \times 10^{-5} \rm \, g \,
cm^{-3}$ would be consistent with standard late K giant or late A
supergiant stars.  However, the properties of the companion star are
significantly affected by the high rate of mass transfer, and stellar
evolution calculations taking into account the mass transfer are
required to properly understand the evolutionary state of the companion
star.

Stellar evolution calculations performed by \citet{li04} for a star
with an initial mass of 15~$M_{\odot}$ orbiting a 1000~$M_{\odot}$
black hole show that the mass transfer rate exceeds $10^{-4} M_{\odot}
\, \rm yr^{-1}$ when the companion evolves through the giant phase at
10~Myr after formation of the binary.  Similar results are obtained by
\citet{portegies04b}.  This mass transfer rate is more than sufficient
to fuel the observed X-ray luminosity of $10^{41} \rm \, erg \,
s^{-1}$.  The lifetime of the giant phase is relatively short, on the
order of  $10^{5}$~years.  This may suggest that the ULX phase is
rather short lived and that systems with very high luminosities, on the
order of $10^{41} \rm \, erg \, s^{-1}$, are rare \citep{kaaret06}.  

As noted in the introduction, X41.4+60 lies near and possibly within
the super star cluster MGG 11.  \citet{mccrady03} estimate that MGG 11
is 7-12~Myr.  Hence, the time required for the companion to evolve to
the giant phase is consistent with the age of the cluster.

A possible mechanism for the formation of intermediate mass black holes
with masses greater than $100 M_{\odot}$ is stellar collisions in the
cores of dense stellar clusters
\citep{portegies99,taniguchi00,miller02}.  \citet{portegies04a} found
that the extremely compact size of MGG 11, the half light radius is
only 1.2~pc, causes massive stars to rapidly sink to the cluster center
via dynamical friction.  The time for sinking is shorter than the
lifetimes of the stars, so a large number of massive stars accumulate
in the cluster center where runaway collisions produce a star of $\sim
1000 M{\odot}$ that collapses to an intermediate mass black hole.  This
process does not occur in less dense clusters.  The coincidence of the
X41.4+60 with a super star cluster with the correct properties for the
production of an intermediate mass black holes favors runaway
collisions as the production mechanism of the ULX. Other models for
intermediate mass black hole formation, such as remnants of the first
generation of stars \citep{madau01} or as the nuclei of captured
satellite galaxies \citep{king04}, would not predict such a
coincidence.

\section*{Acknowledgments}

We thank an anonymous referee for comments which helped improve the
paper.  PK thanks X.-D.\ Li and Simon Portegies Zwart for useful
discussions and the Aspen Center for Physics for its hospitality.  PK
and MS acknowledge partial support from Chandra grant CXC GO5-6087X and
NASA Grant NNG04GP66G.  PK acknowledges support from a University of
Iowa Faculty Scholar Award.


\label{lastpage}


\begin{thebibliography}{}

\bibitem[Allen \& Kronberg(1998)]{ak98} Allen, M.~L., \&  Kronberg,
P.~P.\ 1998, \apj, 502, 218 

\bibitem[Bauer et al.(2001)]{bauer01} Bauer, F.E., Brandt, W.N.,
Sambruna, R.M., Chartas, G., Garmire, G.P., Kapsi, S., Netzer, H.\
2001, ApJ, 122, 182

\bibitem[Begelman(2002)]{begelman02} Begelman, M.C.\ 2002, ApJ, 568,
L97

\bibitem[Bianchi et al.(2002)]{bianchi02} Bianchi, S., Matt, G.,
Fabian, A.C., Iwasawa, K., Nicastro, F.\ 2002, A\&A, 396, 793

\bibitem[Boyd, Smale, Dolan(2001)]{boyd01} Boyd, P.T., Smale, A.P.,
Dolan, J. F.\ 2001, ApJ, 555, 822

\bibitem[Colbert \& Mushotzky(1999)]{colbert99} Colbert, E.J.M.\ \&
Mushotzky, R.F.\ 1999, ApJ, 519, 89

\bibitem[Collin \& Kawaguchi(2004)]{collin04} Collin, S.\ \& Kawaguchi,
T.\ 2004, A\&A, 426, 797

\bibitem[Corbet(2003)]{corbet03} Corbet, R.H.D.\ 2003, ApJ, 595, 1086

\bibitem[David et al.(2005)]{david05} David, L.P., Jones, C., Forman,
W., Murray, S.S.\ 2005, ApJ, 635, 1053

\bibitem[Dewangan et al.(2005)]{dewangan05} Dewangan, G.C.\, Griffiths,
R.E., Choudhury, M., Miyaji, T., Schurch, N.J.\ 2005, ApJ, 635, 198

\bibitem[Dewangan, Titarchuk, \& Griffiths(2006)]{dewangan06} Dewangan,
G.C.\, Titarchuk, L., Griffiths, R.E.\ 2006, ApJL, to appear,
astro-ph/0509646

\bibitem[Elsner et al.(1980)]{elsner80} Elsner, R.F.,  Ghosh, P., 
Darbro, W., Weisskopf, M.C., Sutherland, P.G., Grindlay, J.E.\ 1980,
ApJ, 239, 355

\bibitem[Fiorito \& Titarchuk(2004)]{fiorito04} Fiorito, R.\ \&
Titarchuk, L.\ 2004, ApJ, 614, L113

\bibitem[Frank, King \& Raine(2002)]{frankbook} Frank, J., King, A.R.,
\& Raine, D.J. 2002, Accretion Power in Astrophysics (Cambridge
University Press)

\bibitem[Griffiths et al.(2000)]{griffiths00} Griffiths, R., Ptak, A.,
Feigelson, E.D, Garmire, G., Townsley, L., Brandt, W.N., Sambruna, R.,
Bregman, J.N.\ 2000, Science, 290, 1325

\bibitem[Horne \& Baliunas(1986)]{horne86} Horne, J.H.\ \& Baliunas,
S.L.\ 1986, ApJ, 302, 757

\bibitem[Huang et al.(1994)]{huang94} Huang, Z.~P., Thuan, 
T.~X., Chevalier, R.~A., Condon, J.~J., \& Yin, Q.~F.\ 1994, \apj, 424, 114

\bibitem[Israel \& Stella(1996)]{israel96} Israel, G.L.\ \& Stella, L.\
1996, ApJ, 468, 369

\bibitem[Kaaret et al.(2001)]{kaaret01} Kaaret, P.\ et al.\ 2001,
MNRAS, 321, L29

\bibitem[Kaaret et al.(2003)]{kaaret03} Kaaret, P., Corbel, S.,
Prestwich, A.H., Zezas, A.\ 2003, Science,  299, 365.

\bibitem[Kaaret et al.(2004)]{kaaret04} Kaaret, P., Alonso-Herrero,
A., Gallagher, J.S.\ III, Fabbiano, G., Zezas, A., Rieke, M.J.\ 2004,
MNRAS, 348, L28

\bibitem[Kaaret(2005)]{kaaret05} Kaaret, P.\ 2005, ApJ, 629, 233

\bibitem[Kaaret, Simet, \& Lang(2006)]{kaaret06} Kaaret, P., Simet,
M.G., Lang, C.C.\ 2006, Science, in press

\bibitem[King et al.(2001)]{king01} King, A.R.\ et al.\ 2001, ApJ, 552,
L109

\bibitem[King \& Dehnen(2004)]{king04} King, A.R.\ \& Dehnen, W.\ 2004,
MNRAS, 357, 275

\bibitem[Krauss et al.(2005)]{krauss05} Krauss, M.I., Kilgard, R.E.,
Garcia, M.R., Roberts, T.P., Prestwich, A.H.\ 2005, ApJ, 630, 228

\bibitem[K\"ording et al.(2002)]{kording02} K\"ording, E. Flacke, H.,
\& Markoff, S.\ 2002, A\&A, 382, L13

\bibitem[Kronberg \& Wilkinson(1985)]{kronberg85} Kronberg, P.P.\ \&
Sramek, R.A.\ 1985, Science, 227, 28

\bibitem[Landt et al.(2001)]{landt01} Landt, H.\ et al. 2001, MNRAS,
323, 757

\bibitem[Liu et al.(2002)]{liu02} Liu, J.-F., Bregman, J.N.,
Lloyd-Davies, E., Irwin, J., Espaillat, C., Seitzer, P.\ 2002, ApJ,
621, L17

\bibitem[Li(2004)]{li04} Li, X.-D.\ 2004, ApJ, 616, L119

\bibitem[Liu, Bregman, \& Seitzer(2004)]{liu04} Liu, J.-F., Bregman,
J.N., Seitzer, P.\ 2004, ApJ, 602, 249

\bibitem[Madau \& Rees(2001)]{madau01} Madau, P.\ \& Rees, M.J.\ 2001,
ApJ, 551, L27

\bibitem[Makishima et al.(2000)]{makishima00} Makishima, K.\ et al.\
2000, ApJ, 535, 632

\bibitem[Matsumoto et al.(2001)]{matsumoto01} Matsumoto, H.\ et al.\
2001, ApJ, 547, L25

\bibitem[McClintock \& Remillard(2003)]{mcclintock03} McClintock, J.E.\
\& Remillard. R.A.\ 2003, astro-ph/0306213

\bibitem[McCrady, Gilbert, \& Graham(2003)]{mccrady03} McCrady, N.,
Gilbert, A.M., Graham, J.R.\ 2003, ApJ, 596, 240

\bibitem[McDonald et al.(2002)]{mm02} McDonald, A.~R., Muxlow,
T.~W.~B., Wills, K.~A., Pedlar, A., \& Beswick, R.~J.\ 2002, MNRAS,
334, 912 

\bibitem[Miller \& Hamilton(2002)]{miller02} Miller, M.C.\ \& Hamilton,
D.P.\ 2002, MNRAS, 330, 232

\bibitem[Mirabel \& Rodriguez(1999)]{mirabel99} Mirabel, I.F.\ \&
Rodriguez, L.F.\ 1999, ARA\&A, 37, 409

\bibitem[Misra \& Sriaram(2003)]{misra03} Misra, R.\ \& Sriram, K.\
2003, ApJ, 584, 981

\bibitem[Morgan et al.(1997)]{morgan97} Morgan, E.H., Remillard, R.A.,
Greiner, J.\ 1997, ApJ, 482, 993

\bibitem[Mucciarelli et al.(2006)]{mucciarelli06} Mucciarelli, P.,
Casella, P., Belloni, T., Zampieri, L., Ranalli, P.\ 2006, MNRAS, to
appear, astro-ph/0509796

\bibitem[Ohsuga et al.(2005)]{ohsuga05} Ohsuga, K., Mori, M., Nakamoto,
T., Mineshige, S.\ 2005, ApJ, 628, 368

\bibitem[Portegies Zwart et al.(1999)]{portegies99} Portegies Zwart,
S.F., Makino, J., McMillian, S.L.W., Hut, P.\ 1999, A\&A, 348, 117

\bibitem[Portegies Zwart et al.(2004a)]{portegies04a} Portegies  Zwart,
S.F., Baumgardt, H., Hut, P., Makino, J., McMillan, S. L. W.\ 2004,
Nature 428, 724

\bibitem[Portegies Zwart et al.(2004b)]{portegies04b} Portegies  Zwart,
S.F., Dewi, J., Maccarone, T.\ 2004, MNRAS, 355, 413

\bibitem[Ptak \& Griffiths(1999)]{ptak99} Ptak, A.\ \& Griffiths, R.\
1999, ApJ, 517, L85

\bibitem[Rephaeli \& Gruber(2002)]{rephaeli02} Rephaeli, Y.\ \& Gruber,
D.\ 2002, A\&A, 389, 752

\bibitem[Rodriguez-Rico et al.(2004)]{rvzga04} Rodriguez-Rico, 
C.~A., Viallefond, F., Zhao, J.-H., Goss, W.~M., \& Anantharamaiah, K.~R.\ 
2004, \apj, 616, 783 

\bibitem[Ruszkowski \& Begelman(2003)]{ruszkowski03} Ruszkowski, M.\ \&
Begelman, M.C.\ 2003, ApJ, 586, 384

\bibitem[Taniguchi et al.(2000)]{taniguchi00} Taniguchi, Y., Shioya,
Y., Tsuru, T.G., Ikeuchi, S.\ 2000, PASJ, 52, 533

\bibitem[Smith, Heindl, \& Swank (2002)]{smith02} Smith, D.M., Heindl,
W.A., Swank, J.H. 2002, ApJ, 578, L132

\bibitem[Soria \& Motch(2004)]{soria04} Soria, R.\ \& Motch, C.\ 2004,
A\&A, 422, 915

\bibitem[Strohmayer \& Mushotzky(2003)]{strohmayer03} Strohmayer, T.E.\
\& Mushotzky, R.F.\ 2003, ApJ, 586, L61

\bibitem[Sugiho et al.(2001)]{sugiho01} Sugiho, M.., Kotoku, J.,
Makishima, K., Kubota, A., Mizuno, T., Fukazawa, Y., Tashiro, M.\ 2001,
ApJ, 561, L73

\bibitem[Vaughan(2005)]{vaughan05} Vaughan, S.\ 2005, MNRAS, 431, 391

\bibitem[Warner et al.(2004)]{warner04} Warner, C., Hamann, F.,
Dietrich, M.\ 2004, ApJ, 608, 136

\bibitem[Watarai et al.(2005)]{watarai05} Watarai, K.-Y., Ohsuga, K.,
Takahashi, R., Fukue, J.\ 2005, PASJ, 57, 513

\bibitem[Weisskopf et al.(2004)]{weisskopf04} Weisskopf, M.C., Wu, K.,
Tennant, A.F., Swartz, D.A., Ghosh, K.K.\ 2004, ApJ, 605, 360


\end{thebibliography}
\end{document}